\documentclass[sn-mathphys,Numbered]{sn-jnl}

\usepackage{graphicx}%
\usepackage{multirow}%
\usepackage{amsmath,amssymb,amsfonts}%
\usepackage{amsthm}%
\usepackage{mathrsfs}%
\usepackage[title]{appendix}%
\usepackage{xcolor}%
\usepackage{textcomp}%
\usepackage{manyfoot}%
\usepackage{booktabs}%
\usepackage{algorithm}%
\usepackage{algorithmicx}%
\usepackage{algpseudocode}%
\usepackage{listings}%
\usepackage{dcolumn}
\usepackage[section]{placeins}

\begin{document}

\title[Article Title]{Effects of Cytoskeletal Network Mesh Size on Cargo Transport}


\author[1]{\fnm{Nimisha} \sur{Krishnan}}\email{nkrish01@syr.edu}

\author[2]{\fnm{Niranjan} \sur{Sarpangala}}\email{nsarpangala@ucmerced.edu}

\author[2]{\fnm{Maria} \sur{Gamez}}\email{malfonsogamez@ucmerced.edu}

\author[2]{\fnm{Ajay} \sur{Gopinathan}}\email{agopinathan@ucmerced.edu}

\author*[1]{\fnm{Jennifer L.} \sur{Ross}}\email{jlross@syr.edu}

\affil*[1]{\orgdiv{Physics Department}, \orgname{Syracuse University}, \orgaddress{\street{Crouse Drive}, \city{Syracuse}, \postcode{13104}, \state{NY}, \country{USA}}}

\affil[2]{\orgdiv{Department of Physics}, \orgname{University of California, Merced}, \orgaddress{\street{5200 North Lake Rd}, \city{Merced}, \postcode{95343}, \state{CA}, \country{USA}}}



\abstract{Intracellular transport of cargoes in the cell is essential for the organization and functioning cells, especially those that are large and elongated. The cytoskeletal networks inside large cells can be highly complex, and this cytoskeletal organization can have impacts on the distance and trajectories of travel. Here, we experimentally created microtubule networks with varying mesh sizes and examined the ability of kinesin-driven quantum dot cargoes to traverse the network. Using the experimental data, we deduced parameters for cargo detachment at intersections and away from intersections, allowing us to create an analytical theory for the run length as a function of mesh size. We also used these parameters to perform simulations of cargoes along paths extracted from the experimental networks. We find excellent agreement between the trends in run length, displacement, and trajectory persistence length comparing the experimental  and simulated trajectories. }




\maketitle

\section{Introduction}\label{sec1}

The movement and positioning of large objects inside cells requires energy-using active transport by motor proteins traversing along cytoskeletal filaments \cite{hirokawa1998kinesin}. This process of intracellular transport is responsible for the organization and reorganization that cells need to survive. Intracellular transport is especially important in cells that are long and extended, such as cilia and axons, or particularly crowded and viscous. In mammalian cells, which are differentiated into a myriad of cell types, diffusion of large cellular components is impeded by the complex viscoelastic nature of the cell interior, so active intracellular transport is required. 

Cytoskeletal filaments, microtubules and actin, serve as the tracks for intracellular transport. Microtubules are particularly used for long-distance transport \cite{hirokawa1998kinesin,vale2003molecular}. Prior works have shown that the arrangement of the cytoskeletal filaments can affect the transport properties of single motors and teams of motors attached to cargoes \cite{ross2008cargo,ross2010multiple, ross2008kinesin,ali2016cargo,lombardo2017myosin,lombardo2019myosin,walcott2022modeling,nelson2009random,tabei2013intracellular}. For long-distance transport, the microtubules are arranged in logical parallel bundles to take advantage of kinesin motors that move distally toward the microtubule plus ends and cytoplasmic dynein motors that move inward to the microtubule minus ends \cite{verhey2011kinesin,burute2019cellular}. In other cell types or locations, the cytoskeletal networks are more complicated. For instance, in muscle cells, microtubules create a cross-hatched network creating intersections for organelles and plasma membrane to anchor during large scale extensions and contractions \cite{robison2016detyrosinated,uchida2022cardiomyocyte}. Prior experimental cellular work has demonstrated that the organization of the cytoskeleton can control the association, dissociation, and trajectory of vesicles that can dynamically change in time and space \cite{tabei2013intracellular,conway2014microtubule}.

There are still open questions about how dense, complex, and crowded conditions can regulate, control, and inhibit intracellular transport. In order to probe the parameters of control, we created microtubule networks of varying densities, characterized by the mesh size, the distance between intersections of microtubules. Using these networks we experimentally probed the trajectories of kinesin-laden quantum dots as they traverse the network. The same networks were used as the basis for simulated trajectories for cargoes where the rates of dissociation at intersections and along the filaments were determined from experiments, making for closer comparisons. We also deduced an analytical function for the run length dependence on mesh size to compare to both experiments and simulations. By comparing the experimental, simulated, and theoretical results, we determined the effects of the network mesh size on the motion of kinesin-driven cargoes.

\section{Methods}\label{sec11}

\subsection{Materials and Reagents} 
Unless otherwise stated, all reagents were purchased from ThermoFisher. 

\subsubsection{Microtubule preparation}
Lyophilized 488-tubulin and unlabeled-tubulin were purchased from Cytoskeleton. Tubulin was resuspended in PEM-80 (80 mM PIPES pH 6.9, 1 mM MgCl$_2$ and 1 mM EGTA) to a final concentration of 5 mg/ml. Labeled tubulin was added to unlabeled tubulin at a 1:10 ratio. To polymerize the tubulin, we added GTP to a final concentration of 1 mM  and incubated for 20 min at 34$^o$C. Finally, we added 20 $\mu$M Taxol to stabilize the polymerized microtubules and incubated at 34$^o$C for 20 min to equilibrate the Taxol. Microtubules were kept on the bench and further dilutions of microtubules required 20 $\mu$M Taxol to keep the filaments stabilized. 

\subsubsection{Kinesin preparation}
Kinesin motors were expressed and purified from the pWC2 plasmid available at AddGene to create a protein with a kinesin-1 motor truncated at amino acid 401, a BCCP tag to allow biotinylation during expression in bacteria, and a 6x-his tag to purify using a nickel affinity column. Kinesin was purified using standard protocols previously described \cite{gilbert1993expression,pierce199814,conway2012motor}. Briefly, the plasmid containing the kinesin construct was transfected into BL21 cells (New England Biolabs) and bacteria were selected using ampicillin in the media. Overnight cultures that included biotin in the media were pelleted and the bacteria was lysed using sonication and chemical lysis. The supernatent was separated from the bacteria debris using centrifugation and then incubated with nickel beads to bind the 6x-his tagged protein. Kinesin was eluted using imidizole and fractions with kinesin were desalted to remove excess imidizole. Kinesin was aliquoted and snap frozen using liquid nitrogen and stored at -80$^o$C. 

\subsubsection{Microtubule network preparation}
\begin{figure}[hbt!]
\centering
\includegraphics[width=0.7\textwidth]{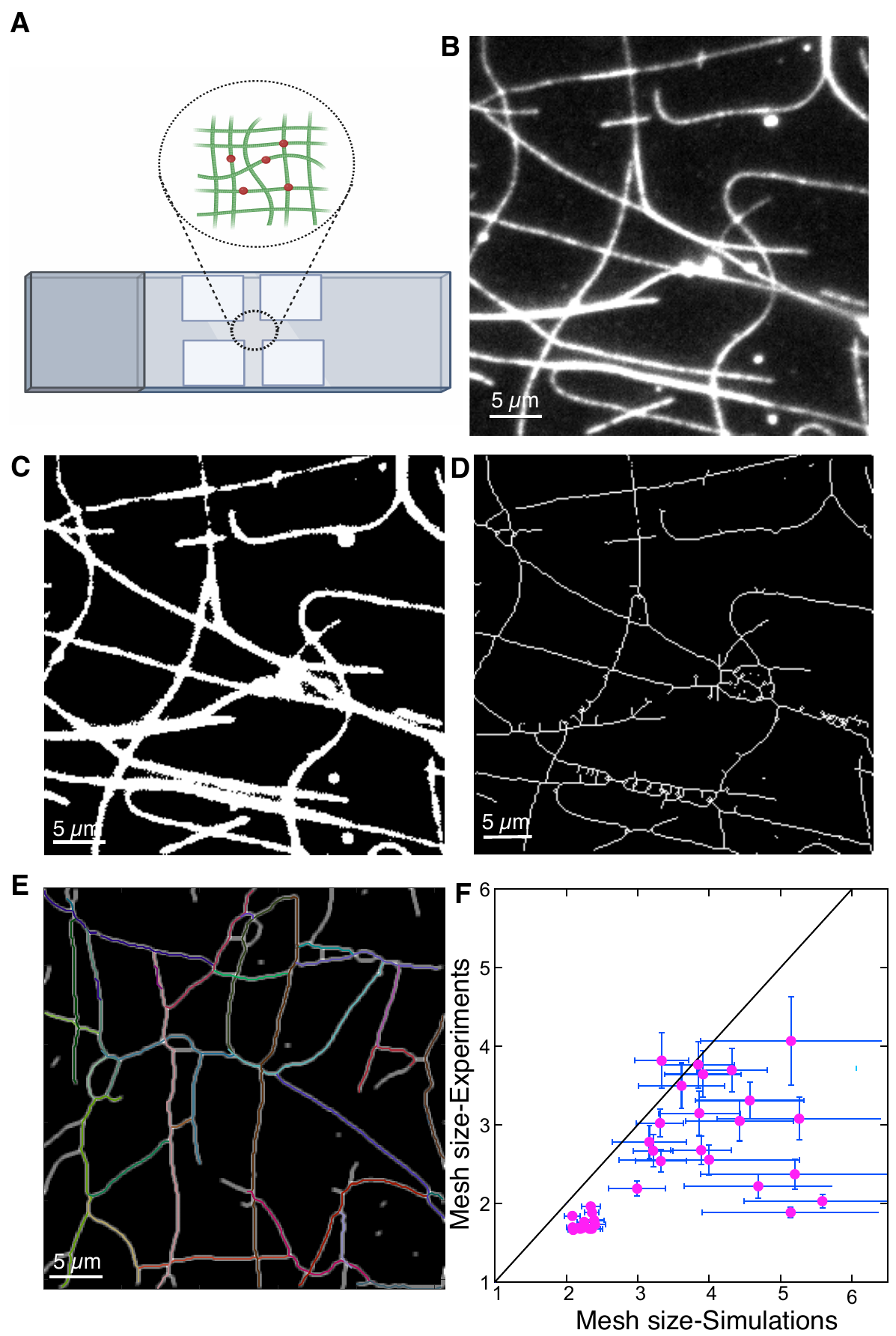}
\caption{\label{fig:network} Microtubule network creation and analysis. (A) Cartoon schematic of crossed flow path sample chamber and microtubule network. (B) Example image of microtubule channel for a network created in a crossed channel chamber. (C) Network image from panel (B) binarized to make a black and white image to be used for mesh analysis. (D) Skeletonization of network used to automatically detect intersections and branch lengths. (E) Extracted network used to perform simulations of motors on networks with the same organization as experiments. (F) Comparison between the mesh size measured from ImageJ and extracted using MatLab. Not all networks used in experiments were extracted and used for simulations. For all images, the scale bar is 5 $\mu$m. }
\end{figure}

Microtubule networks of varying filament mesh density were made by flow-aligning microtubules in a crossed-path flow chamber as previously described \cite{ross2008kinesin}. We made the chamber by adhering four square pieces of double-sided tape on a glass slide such that it made a crossed flow path (Fig. \ref{fig:network}A). The slide was bound to a silanized cover glass treated with hydrophobic silane, PlusOne Repel silane (Cytiva) as previously described \cite{dixit2010studying}. 

To create the sample, the following reagents were flowed into the chamber. First, we flowed 15 $\mu$l of 10$\%$ $\alpha$-tubulin antibody (YL1/2) into the chamber and incubated for 5 minutes. This surface layer provided a specific interaction to the microtubules and helped to elevate them above the polymer surface coating, which was added next. The polymer surface was made by adding 10 $\mu$l of 5$\%$ Pluronic F-127 block copolymer from both directions of the flow chamber and incubating for 5 min. The pluronic blocks the surface from other proteins non-specifically binding. Next we washed the chamber with wash buffer (90 $\mu$l PEM-80, 10 $\mu$l of 0.5$\%$ Pluronic F-127). Now that our surface was well coated and blocked, we flowed 10 $\mu$l of polymerized microtubules diluted to 0.5 mg/ml tubulin concentration. We flowed from the x-direction, incubated for 2 minutes, washed with wash buffer, and incubated for another 3 minutes. We repeated the same process in the y-direction. The chamber was imaged to ensure microtubule networks were bound to the surface and at the densities needed. 

\subsubsection{Quantum-dot cargo preparation}
\begin{figure}[hbt!]
    \centering
    \includegraphics[width=0.7\textwidth]{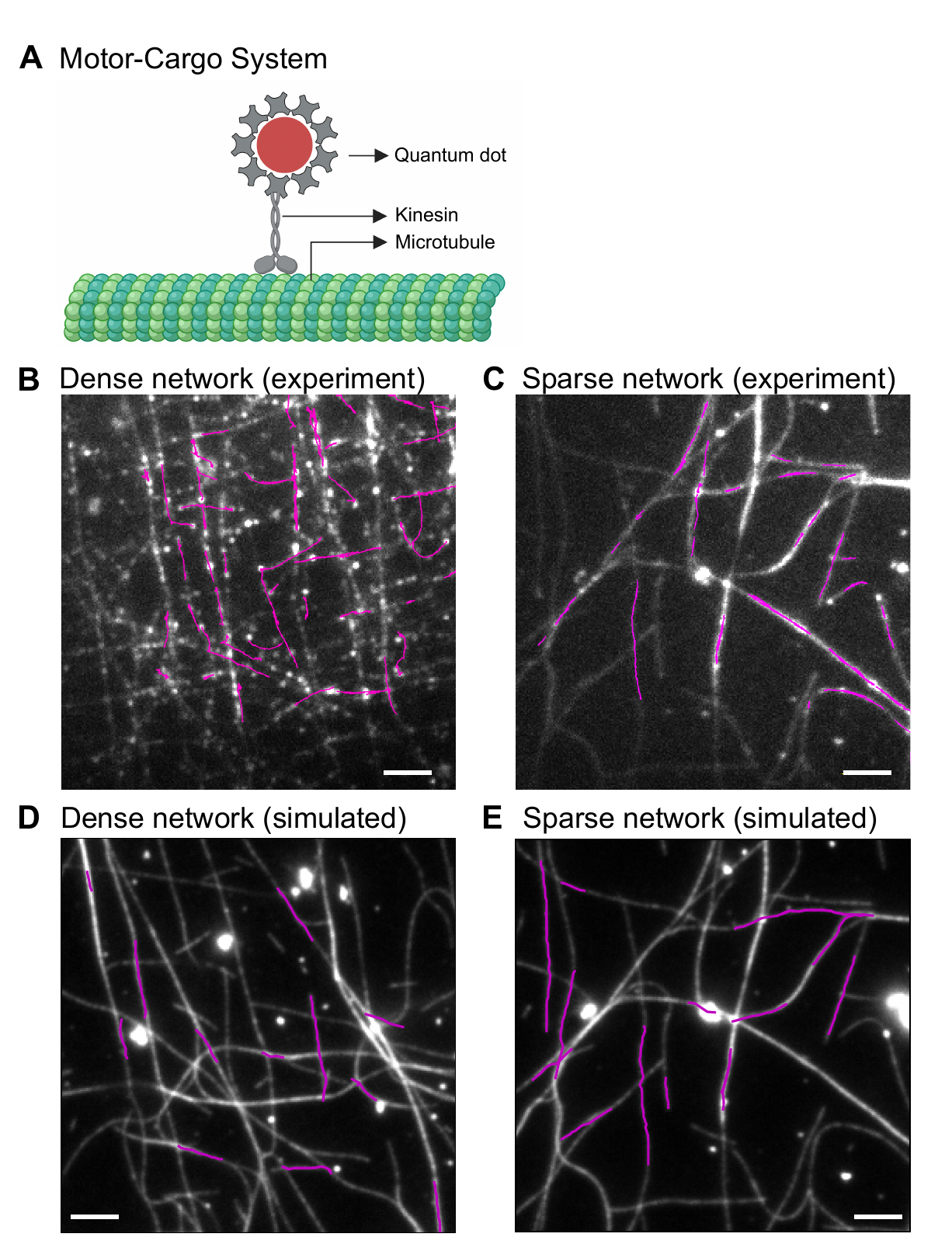}
    \caption{\label{fig:qdotcargo} Quantum dot cargo methods and trajectory analysis. (A) Cartoon schematic of quantum dot cargo attached to a kinesin motor that can walk along microtubules. (B) Example cargo trajectories (magenta) are displayed on a dense microtubule network (white) with a small mesh size. Only a subset of total trajectories are shown. (C) Example cargo trajectories (magenta) along a sparse microtubule network (white) with a large mesh size. (D) Example simulated trajectories (magenta) along dense extracted network (white) with a small mesh size. (E) Example simulated trajectories (magenta) along sparse extracted network (white) with a large mesh size. For all images, the scale bar is 5 $\mu$m.}
\end{figure}

Quantum dot cargoes were made by mixing streptavidin-labeled quantum dots (ThermoFisher) with biotinylated kinesin at a ratio of 1:2 and incubated for one hour on ice (Fig. \ref{fig:qdotcargo}). These cargoes were then diluted by 30 times in PEM-80 to be used in microtubule networks of varying densities. The final step of sample chamber preparation was to add the kinesin cargo sample to the chamber that has been examined on the microscope. The final flow through contains quantum dots diluted to 1:30 in PEM-80 with 2 mM ATP, 66 mM DTT, and an oxygen scavenging system, which was 0.66 mg/ml glucose oxidase, 1.5$\%$ final dilution of aqueous catalyse (Sigma catalogue number C30), and 20 mg/ml glucose in PEM-80. The microtubule network in the crossed part of the flow chamber had varying densities allowing us to take data in several locations within the same chamber (Fig. \ref{fig:network}). If needed, kinesin cargo sample was replenished into the same sample chamber to replace the ATP and oxygen scavenging species that degraded during the assay and allow longer imaging.

\subsubsection{Microscopy imaging}
Image data was captured with a Nikon Ti-E microscope using epi-fluorescence and total internal reflection fluorescence (TIRF) microscopy as previously described \cite{conway2012motor,ross2010multiple,ross2008kinesin}. Microtubules were imaged in epi-fluorescence in the green fluorescence channel using a Hg-Xe illumination source with 480 $\pm$ 25 nm excitation filter, a 500 nm long pass filter, and a 525 $\pm$ 55 nm emission filter (Chroma). The illumination for the TIRF system was a custom-built laser system using a 647 nm solid state laser brought into the back of the 60x, 1.49 NA objective, as previously described \cite{ross2010multiple}. The filter set had no excitation filter, a 640 nm long pass for the dichroic, and a 680 $\pm$ 50 nm emission filter (Chroma). All images were made using an IXON electron-multiplier CCD camera (Andor) with a pixel size of 160 nm. The laser and camera systems were controlled through Nikon Elements software and images were recorded to RAM and saved a .nd2 files as uncompressed tif stacks and metadata. The time series data sets of quantum dot cargoes were taken for 2 mins with 1 s in between frames with an exposure time of 100 ms. 

\subsection{Quantitative image analysis}
\subsubsection{Network mesh size characterization}
The control parameter for these studies was the network mesh size, which was denoted as the distance between neighboring intersections of the microtubule network. We noticed that the mesh size could change significantly over the imaging region of our camera, which was 82 $\mu$m on a side. In order to have the entire region have a similar mesh size, we divided each image into quarters for analysis of both the mesh size and the trajectories (see below). This gave more consistent network mesh sizes over an area of 41x41 $\mu m^2$. The same networks were extracted and used as the basis for simulated trajectories (see below). 

We quantified the distance between intersections using the FIJI/ImageJ AnalyzeSkeleton (2D/3D) plugin \cite{lee1994building}. First, we smoothed the images to remove fluctuations in the background caused by shot noise. Next, we performed background subtraction on the images to remove global intensity variations due to the imaging. We converted the image into a binary image by using the auto threshold function to make the microtubules white on a black background (Fig. \ref{fig:network}C). This helps in distinguishing signal (microtubules) from background. We then skeletonized the image using the binary/skeleton command in FIJI/ImageJ (Fig. \ref{fig:network}D). Finally, we applied the AnalyzeSkeleton (2D/3D) plugin with prune ends and prune cycle shortest branch enabled. The data reported were the largest shortest path, detailed info, and labeled skeletons. The resulting data displayed the branch information, which was saved as a text file. We used the branch length given in microns as the data for the mesh size. The statistics of the data were calculated and the distribution was normal, so the mean and the median were the same. The error bars reported are the standard error of the mean. The number of branches analyzed are given in Appendix table \ref{table:NvaluesMeshes}. 

\subsubsection{Tracking and transport analysis}
We used Fiji/ImageJ tracking plugin Trackmate \cite{tinevez2017trackmate} to track quantum dot trajectories. Within Trackmate, we used a setting of 6 pixels for the diameter of the objects and we allowed a gap in time of 2 frames, so that more than 2 frames without detecting the object nearby resulted in terminating the measurement. We also used a minimum cut off run length of 3 pixels (160 nm/pixel) and 3 frames (1 s/frame). The localization was allowed to be sub-pixel. Each of the tracks were manually checked against the movie to ensure the trajectory tracked was reasonable. The x,y position data over time were used as the trajectories for further analysis. Example trajectories on different networks are show in figure \ref{fig:qdotcargo}. The number of tracks analyzed for each network is given in Appendix table \ref{table:NvaluesMeshes}.

The run length of a trajectory was determined as the contour length of the trajectory where the absolute value of all the displacements were summed. The displacement of a trajectory was defined as the end-to-end distance of the trajectory. The instantaneous speed was calculated as the positive displacement between two frames divided by the time between frames. The average speed was determined as the run length divided by the total time the quantum dot was associated to the network. For all data types, the data was averaged and the standard deviation or standard error of the mean was used as error bars. The number of tracked trajectories used for each network is given in Appendix table \ref{table:NvaluesMeshes}.

To quantify the characteristic persistence length of the trajectory, we calculated the mean squared displacement (MSD) and plotted it versus the contour length of the trajectory. The MSD was calculated by measuring the displacement for all points along the trajectory at a specific lag time (time between frames). For each lag time, the displacement values were squared and averaged. This is performed for all the lag times that have more than 5 data points and plotted for only the first 80 points. Once we calculate the MSD as a function of lag time, we also determined the run length (total contour length from initial time) of the trajectory for all time. The run length from zero time is plotted as the x-coordinate and the MSD as a function of lag time is plotted on the y-coordinate. The MSD was fit to a worm-like chain model with this equation: 
\begin{equation}\label{eq:WLC}
    MSD(L_c)=2L_p L_c \left[ 1 - \frac{L_p}{L_c} \left( 1 - \exp(-L_c/L_p \right) \right]
\end{equation} 
where $MSD(L_c)$ is the mean squared displacement as a function of the contour length, $L_c$, and $L_p$ is the persistence length, which is a fit parameter.

\subsection{Network extraction and characterization}
We used the Matlab tool, FIRE \cite{Andrew2008} on skeletonized images (Fig. \ref{fig:network}D) to extract corresponding networks (numerical matrices of filaments and vertices) for cargo transport simulations. We filtered out any filaments that were too short to be a real filament (typical cut-off length is 0.48 $\mu$ m). We then characterized the network by quantifying the mesh size, as the mean of the distances between filament intersections and the persistence length of filaments. To obtain the persistence length, we restructured the network data so as to represent filaments as trajectories on a 2D cartesian plane. Then we measured the MSD of these paths as a function of path length, and fit the worm-like chain model (Eq. \ref{eq:WLC}) as described above.

\subsection{Simulations on extracted network}
The computational model for cargo transport on the network was similar to previous works \cite{Ando2015, Bryan2019}. A cargo of radius $r$ was initialized at a random point in the 2-dimensional box containing the extracted network. It was allowed to diffusively search and bind to a filament. The simulation time starts from the time when the cargo bound to the filament. After binding, the cargo ballistically moved toward one of the filament ends, \textit{i.e.} in a time step $\Delta t$, the cargo position moved by $v\Delta t$ towards the next vertex, where $v$ is the velocity of cargo. The polarity of the filament was chosen randomly when the cargo bound and was fixed throughout the given cargo run. As a cargo walked ballistically on the filament, it could stochastically detach anywhere along the path with a rate, $k_{off}=v\tilde{k}_d$, where $\tilde{k}_d$ was the detachment rate per unit length along a filament. In addition to stochastic detachments a the cargo walks along a filament, it also could detach with a fixed probability, $P_d$ at filament intersections. A cargo run stopped when it detached from the filament. Cargo was assumed to interact with the intersection when it was closer than one cargo radius ($r$) to one of the filament intersections. The value of the off rate, $\tilde{k}_d$ and detachment probability at intersections, $P_d$, were determined from the analysis of run lengths from manual tracking of experimental videos (see Sec 3.2). Finally, the cargo was assumed to detach from filaments when it reached one of the filament ends. In our model, we simulated only cargo transport along the filament and considered only ballistic motion for analysis to compare with experimental tracks. This was different from previous models where diffusive transport of cargo in the cytoplasm was considered \cite{Ando2015, Bryan2019} in addition to the ballistic phase on filaments. Since our goal was to analyze the impact of network features on transport, which is independent of time, we decided to work with quantities that do not depend on time. Instead, we examined run length, displacement, and MSD as a function of path length to compare with experiments. Thus $v$ in our model was a free parameter that was chosen depending on computational convenience. We performed N=1000 cargo runs for each network.

\section{Results and Discussion}\label{sec2}
\subsection{Network mesh size}
We created artificial cargoes from quantum dots decorated with kinesin-1 motors added to microtubule networks of various mesh sizes. The mesh size of the networks and a variety of transport parameters were quantified to determine how the microtubule network density affected the mobility of kinesin cargoes. The average mesh sizes measured from experimental images using ImageJ ranged from 1.5 $\mu$m to 5 $\mu$m between intersections (Fig. \ref{fig:network}F). The experimental networks were extracted to be used for simulations. When the mesh sizes were characterized from simulations using MatLab, they were 37$\%$ higher than the experimental characterization on 97$\%$ of the networks (Fig. \ref{fig:network}F). It is possible that in the extraction, we lost small intersections that were actually there in experiments, resulting in a larger mesh size in simulations. This could result in altered quantitative results from simulations compared to experiments. We expected that the trends should be the same, if the correct underlying mechanisms were being simulated, which we test below. For plotting of experimental and simulated data, we chose to use the characteristic mesh sizes determined from experimental images using ImageJ in all figures.  

\subsection{Run length}
\begin{figure}[hbt!]
\centering
\includegraphics[width=0.8\textwidth]{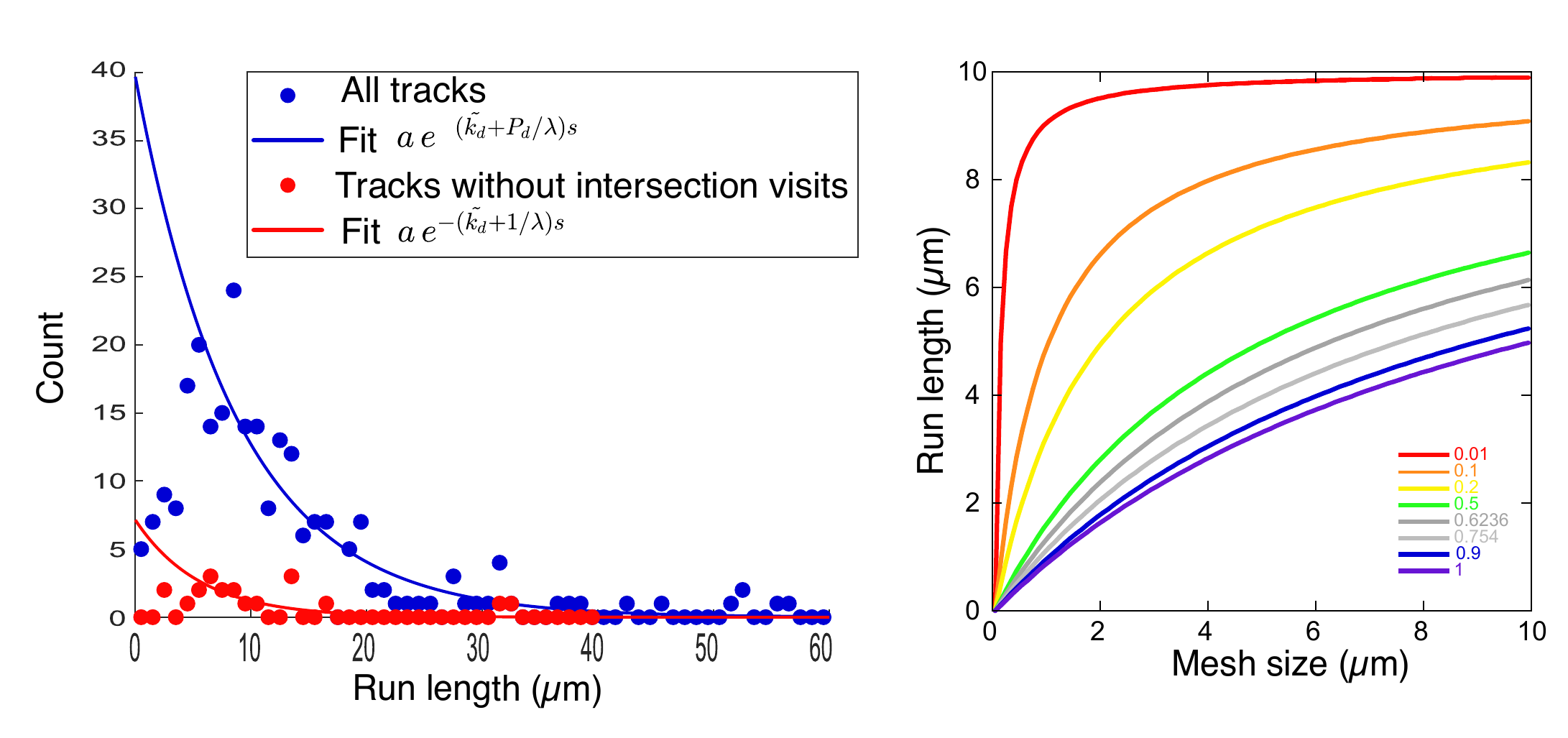}
\caption{\label{fig:runlength2} Determination of simulation parameters and analytical theory. (A) Distribution of run lengths of all tracks (blue circles, N=246) fit to exponential decay (blue line). Distribution of run lengths of tracks that did not visit an intersection (red circles, N=20) fit to exponential decay (red line). Short run lengths were excluded from the fit because manual tracking had a systematic bias against short run lengths. (B) Analytical theory for the average run length as a function of mesh size (Eq. \ref{eq:analytical}) for various values of $P_d$. The legend indicates the value of $P_d$ plotted with the two values from the histograms, 0.62 and 0.75 in gray.}
\end{figure}

The run length of a quantum dot cargo is the total distance, or contour length, traveled along the microtubules of the network before detaching. Experimental data was collected and long trajectories were manually tracked, regardless of network mesh size. The run length, $s$, was determined from the manually tracked trajectories and the data was binned with either 8 $\mu$m or 1 $\mu$m bins (Fig. \ref{fig:runlength2}A, blue). We deduced the two parameters that controlled run length via dissociation, namely the off rate for cargoes between intersections,  $k_{off}$, and the probability of detaching at the intersections, $P_d$, by fitting the run length histograms to exponential decay functions of the form: 
$y = a e^{-(\tilde{k_d}+P_d/\lambda)s}$,
where $\lambda$ is the mean mesh size for all extracted networks, $a$ is an arbitrary normalization parameter, and $\tilde{k_d}$ is the detachment rate per unit length. The parameter $\tilde{k_d}$ is equivalent to the inverse of the ``natural'' run length for the cargoes in the absence of intersections and is equal to the off rate between intersections divided by the cargo velocity: $k_{off}/v$  (Fig. \ref{fig:runlength2}A, blue). 

Because we have two unknown parameters, we also need to use a second set of data to deduce the off rate between intersections. Using only the subset of tracks that never visited an intersection during their trajectories, the run lengths were again binned and fit to $y = a e^{-(\tilde{k_d}+1/\lambda)s}$ (Fig. \ref{fig:runlength2}A, red). The two histograms had fit parameters of $A=\tilde{k_d}+P_d/\lambda$ and $B=\tilde{k_d}+1/\lambda$, which were used to deduce $\tilde{k_d}$ and $P_d$ (Table \ref{table:simulationparameters}). 

The histogram bin size had an effect on the fitting parameters and hence the values of $k_{off}$ and $P_d$ used for simulations. Histograms with 8 $\mu$m bins resulted in a $P_d$ value of 0.75 and histograms with 1 $\mu$m bins resulted in a $P_d$ value of 0.62. Assuming the distribution of run lengths has this form: 
$a e^{-(\tilde{k_d}+P_d/\lambda)s}$,
then the average run length, $\langle s \rangle$, for a network with a mesh size, $\lambda$, should be given by finding the average of this expression given by: 
\begin{equation}\label{eq:analytical}
\langle s \rangle = 1/(\tilde{k_d}+P_d/\lambda).
\end{equation}
Thus the dependence of the average run length ($\langle s \rangle$) with mesh size ($\lambda$) is non-linear and saturates to $1/\tilde{k_d}$ for large mesh sizes (Fig. \ref{fig:runlength2}B). Given the uncertainty of the estimation of $P_d$, we can examine the sensitivity of this analytical expression to the value of $P_d$ (Fig. \ref{fig:runlength2}B). For small $P_d$, the run length saturates to the natural run length for cargoes on a single microtubule. When $P_d$ approaches 1, the run lengths are depressed and only reach the natural run length at higher mesh sizes (Fig. \ref{fig:runlength2}B). This theoretical equation is general in that it allows us to compute run lengths at any values of motor off rate, mesh size, and detachment probability at intersections. This equation will also allow future experiments to compute the detachment probability at intersections without having to manually filter trajectories that encounter intersections. For the simulations, we used a value of $P_d$ equal to 0.75 and a $\tilde{k_d}$ equal to 0.1.

\begin{figure}[hbt]
\centering
\includegraphics[width=0.55\textwidth]{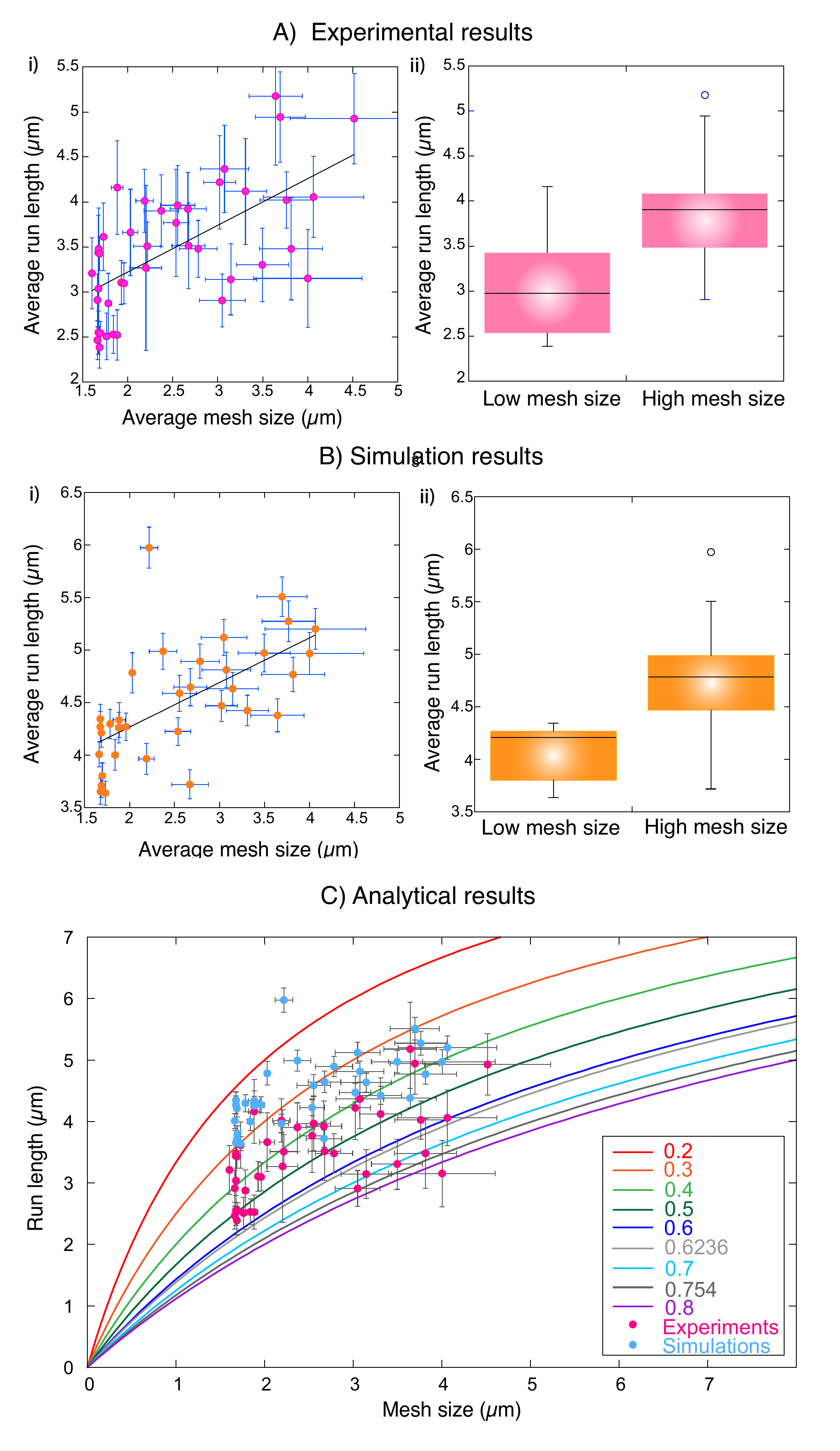}
\caption{\label{fig:runlength} Total run lengths for experimental and simulated trajectories. (A) Experimental results. (i) Plot showing the average run length ($\mu$m) for each network of a given average mesh size ($\mu$m). Best fit slope is 0.5 $\pm$ 0.1. (ii) Comparison of the distribution of the average run lengths for networks with low mesh size less and high mesh size. (B) Simulation results. (i) Average run lengths of simulated trajectories on the same networks. Each data point represents average over N=1000 cargo runs, error bars represent standard error of the mean. Best fit slope is 0.4 $\pm$ 0.1. (ii) Comparison of the distribution of the average run lengths for low mesh size and high mesh size. All fit parameters for data are given in Appendix table \ref{table:FitParameters}. (C) Comparison of experimental run lengths (magenta circles) and simulated run lengths (blue circles) as a function of network mesh size with the analytical theory with various values of $P_d$, given in the legend. }
\end{figure}

We compared our analytical results to both our experimental quantification of run length and simulations of the run length on our microtubule networks. For the experimental data, we observed that as the mesh size increased, the average run length also increased (Fig. \ref{fig:runlength}Ai,ii). We assigned an artificial cut-off between low mesh size (less than 2 $\mu m$) and high mesh size (larger than 2 $\mu m$). Using this cut-off, we found that the median run length for low mesh size was 3.0 $\pm$ 0.1 $\mu m/s$ and the high mesh size was 3.9 $\pm$ 0.1 $\mu m/s$ (Fig. \ref{fig:runlength}Aii). 

There was a distinct difference in the statistics for networks with mesh sizes above and below this cut-off (Fig. \ref{fig:runlength}Aii). Indeed, performing the Kolmogorov-Smirnoff statistical test (KS Test), we found the probability was p = 0.0003 or 0.03\% that the small and large mesh size results were the same (Fig. \ref{fig:runlength}Aii). Thus, we concluded that the threshold at a mesh size of 2 $\mu m$ is a reasonable cut-off between low and high mesh sizes for further comparisons. The experimental run length results implied that quantum dots cover longer distances when the microtubule tracks were more open, with fewer intersections. This result makes sense, since smaller mesh sizes should have more intersections, and kinesin has been shown to have a high probability of terminating a run when contacting a microtubule intersection \cite{ross2008kinesin}.  

Using the deduced probabilities for dissociation between or at intersections (Fig. \ref{fig:runlength2}A), we were able to simulate trajectories on different networks and quantify the run lengths as a function of mesh size (Fig. \ref{fig:runlength}B). The simulated run lengths increased with increasing mesh size just like the experimental data with a similar slope (Fig. \ref{fig:runlength}Bi). Using the same mesh size cut-off, we compared the simulated run lengths to find the low mesh size median run length was 4.2 $\pm$ 0.8 and the high mesh size median run length was 4.8 $\pm$ 0.1. These results were significantly different with a probability of 0.0015\% that they are the same distribution using the KS Test.

We can compare the experimental and simulated trajectory run lengths to the analytical expression for average run length as a function of mesh size (Eq. \ref{eq:analytical}, Fig. \ref{fig:runlength}C). Plotting all together, it is clear that the simulations have systematically longer run lengths than the experimental results. Given the sensitivity to the value of $P_d$, it is possible that adjusting this parameter or the natural run length, $1/\tilde{k_d}$, could cause the difference. Examining the data with multiple values of $P_d$ plotted, we estimate that the probability of detaching at an intersection is between 0.3 and 0.75 for the experimental data and between 0.2 and 0.5 for the simulation data (Fig. \ref{fig:runlength}C). This is a not surprising considering that the extracted networks have larger mesh sizes compared to experiments (Fig. \ref{fig:network}). Thus, for the same values of $P_d$ and $1/\tilde{k_d}$, the number of intersections encountered is smaller, and the run lengths will thus be larger. 

Another difference between simulations and experiment could results from track switching at intersections. In our simulations, we did not include track switching as an option at intersections (Table \ref{table:simulationparameters}) because switching has been shown to be infrequent for kinesin cargoes with one or two motors \cite{ross2008kinesin}. Further, in our manual tracking of long trajectories, we only observed switching at intersections with about 5\% probability, which matches prior reports for single GFP-kinesin \cite{ross2008kinesin}. In order to check if 5\% switching probability at intersections could alter the results, we simulated trajectories in the networks including this probability. We found that this small switching probability has no effect on the run lengths we observe in simulations (Appendix Fig. \ref{fig:simulationfig}1), justifying our choice to not include it in the simulations. Further, this also implies that there is likely on average one active motor per quantum dot, although there is a small probability that some quantum dots have two motors. 

\subsection{Instantaneous and average speeds} 
\begin{figure}[hbt!]
\centering
\includegraphics[width=0.7\textwidth]{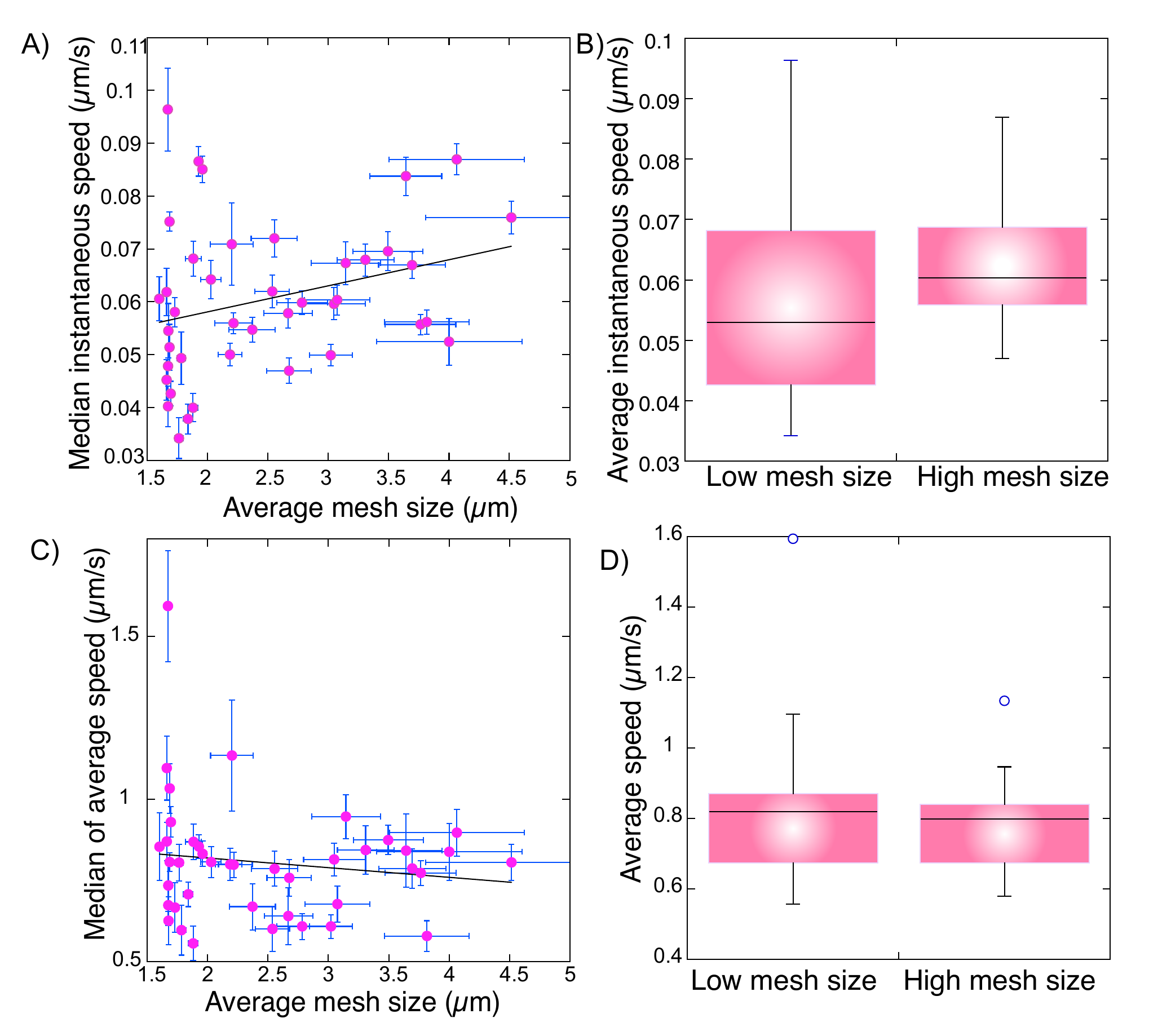}
\caption{\label{fig:speed} Cargo speeds were independent of mesh size. (A) Plot showing the median instantaneous speed ($\mu$m/s) for each network of a given average mesh size ($\mu$m). The best fit slope was 0.005 $\pm$ 0.003.  (B) Comparison of the distribution of the median instantaneous speed for low and high mesh sizes, which were statistically the same. (C) Plot showing the median average speed ($\mu$m/s) for each network of a given average mesh size ($\mu$m) with the best fit slope of -0.03 $\pm$ 0.03. (D) Comparison of the distribution of the median average speed for low and high mesh sizes, which were statistically the same. All fit parameters for data are given in Appendix table \ref{table:FitParameters}. }
\end{figure}

We used metrics that were independent of time to allow us to compare between the experiment and simulation data. In our model, the speed $v$ was a free parameter that was chosen depending on computational convenience. Given these assumptions, we would expect that the average and instantaneous velocities should be independent of mesh size. As a rationale check, we calculated the instantaneous speed of the quantum dot cargoes between two time points. We found that there was a shallow trend in the median of the instantaneous speed as a function of average mesh size (Fig. \ref{fig:speed}A). Using the same threshold at a mesh size of 2 $\mu m$, we found that the two distributions have different medians of 0.052 $\pm$ 0.004 $\mu m/s$ for the low mesh size and 0.060 $\pm$ 0.002 $\mu m/s$ for the high mesh size (Fig. \ref{fig:speed}B). The distribution of instantaneous speeds was wider with a standard deviation of 0.018 $\pm$ 0.004 compared to the standard deviation of the high mesh size data, which had a standard deviation 0.010 $\pm$ 0.002 (Fig. \ref{fig:speed}B). This was likely because, at a smaller mesh size, there was a higher probability of a cargo encountering an intersection between two frames and affecting the instantaneous speed. Comparing the data with the KS Test, we found that the distribution in instantaneous velocities for the low and high mesh sizes were not statistically different, with a 5.6\% probability that they are the same. 

We also quantified the average velocity of the cargo trajectories, given by the total run length (Fig. \ref{fig:runlength}) over the total association time. Since the network intersections reduced both the association time and the run length by the same mechanism, specifically causing the dissociation of the cargoes, we expected both parameters to decrease similarly. Indeed, we found that the average speed of the cargoes was unaffected by mesh size, as expected (Fig. \ref{fig:speed}C). The average speed distributions had medians of 0.82 $\pm$ 0.06 for low mesh and 0.80 $\pm$ 0.03 for high mesh size (Fig. \ref{fig:speed}D). The probability that these two distributions are the same is 39\% using the KS test (Fig. \ref{fig:speed}D). These checks indicated that our model assumptions about velocity were reasonable.  

\subsection{Displacement and tortuosity} 
\begin{figure}[hbt!]
\centering
\includegraphics[width=0.75\textwidth]{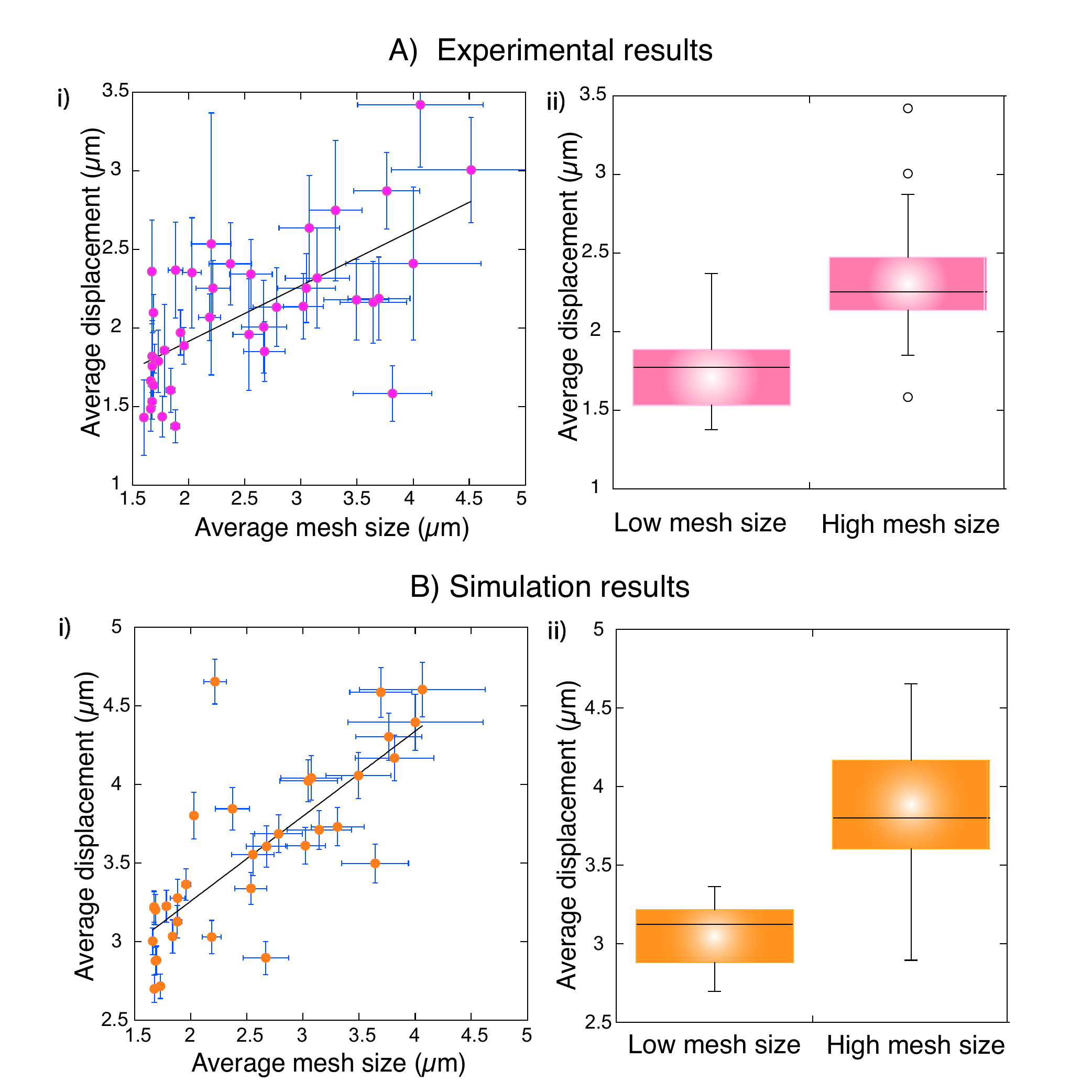}
\caption{\label{fig:displacement} Total displacement for experimental and simulated trajectories.  (A) Displacement of experimental trajectories. (i) Plot showing the average displacement ($\mu$m) against average mesh size ($\mu$m). The best fit slope is 0.35 $\pm$ 0.06. (ii) Comparison of the distribution of average displacements for low mesh size and high mesh size. (B) Displacement of simulated trajectories. (i) Plot showing the average displacement ($\mu$m) with average mesh size ($\mu$m). The best-fit slope is 0.5 $\pm$ 0.1. (ii) Comparison of the distribution of average displacements for low mesh size and high mesh size. All fit parameters for data are given in Appendix table \ref{table:FitParameters}. }
\end{figure}

The displacement is the end-to-end length of the cargo's trajectory. We quantified the displacement for experimentally measured trajectories and found that it increased with the mesh size (Fig. \ref{fig:displacement}Ai). The median displacement for low mesh size was 1.77 $\pm$ 0.07 and for high mesh size was 2.26 $\pm$ 0.08 (Fig. \ref{fig:displacement}Aii). The probability that the distributions in displacement were the same was 0.01\% using the KS Test. 

The simulated trajectories also showed the same trend in displacement, increasing with mesh size (Fig. \ref{fig:displacement}Bi). The median displacement for low mesh size was 3.12 $\pm$ 0.06 and for high mesh size was 3.80 $\pm$ 0.01 (Fig. \ref{fig:displacement}Bii). The probability that the distributions in displacement were the same was 0.0015\% for high and low mesh sizes using the KS Test.  Although the trends are similar, the absolute numbers for the simulated displacements were higher than the measured displacements for the same networks. Like the run length, the displacements of simulated trajectories were not affected by a 5\% probability to switch microtubules at intersections (Appendix Fig. \ref{fig:simulationfig}1).

We can determine the tortuosity of the trajectories by dividing the contour length by the displacement. This is a parameter used for examining the flow of material through porous media, and can be used to characterize the mobility. We find that the average tortuosity for low mesh size networks is 1.7 $\pm$ 0.1 and for high mesh size is 1.6 $\pm$ 0.1, which are the same. Thus, the run length and displacement are being rescaled within the network by the same process, most likely the presence of the intersections.

\subsubsection{Mean square displacement}
\begin{figure}[hbt!]
\centering
\includegraphics[width=1\textwidth]{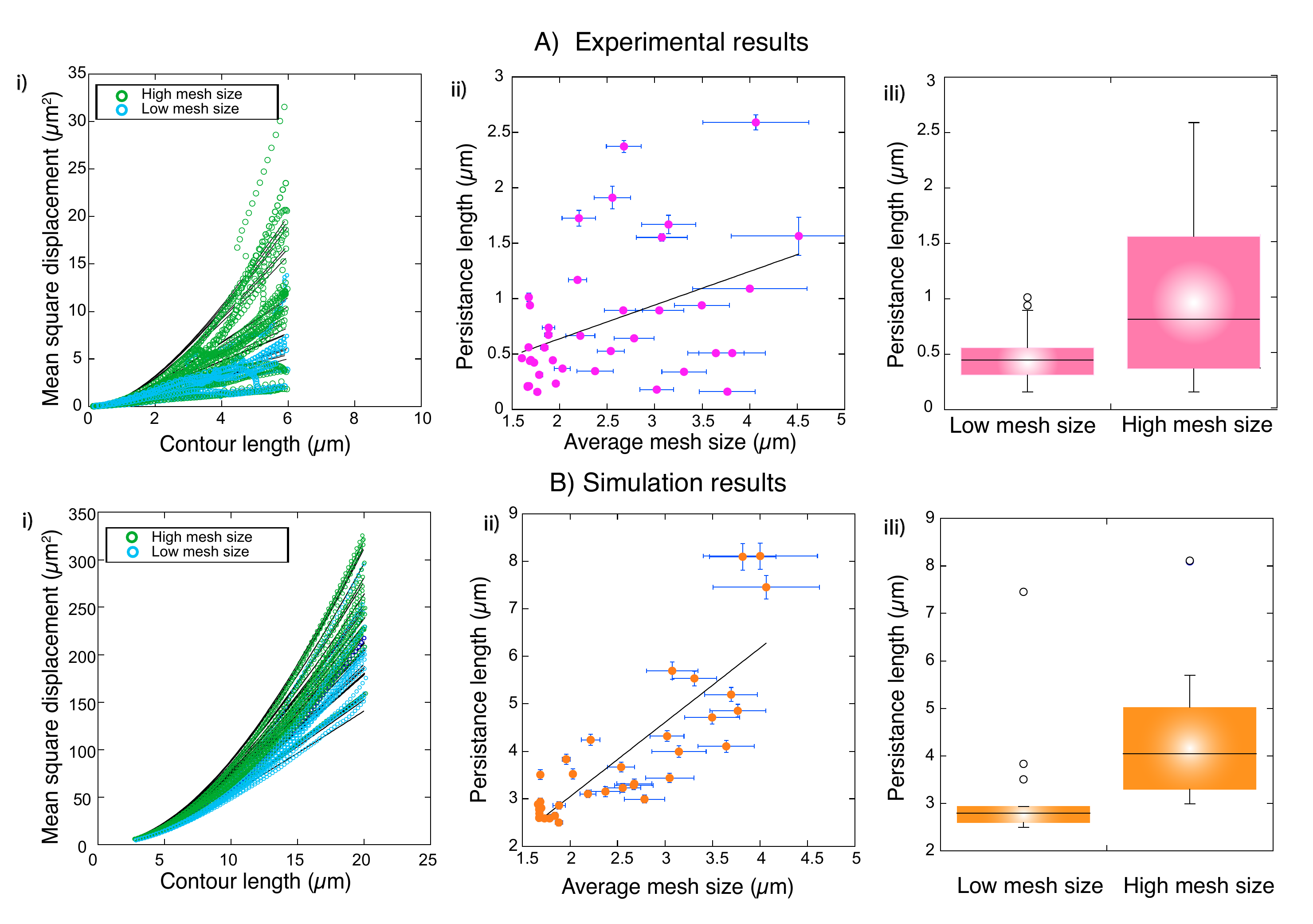}
\caption{\label{fig:MSD} Mean squared displacement and persistence length of trajectories. (A) Mean squared displacement for experimental data (i) All experimental data sets from low mesh size (cyan circles) and high mesh size (green circles) plotted together. These data were fit to the worm-like chain model (Eq. \ref{eq:WLC}) to find the persistence length. (ii) Persistence length ($\mu$m) plotted against mesh size ($\mu$m). The best fit slope is 0.2 $\pm$ 0.1 (iii) Comparison of median of persistence lengths for low mesh size and high mesh size. (B) Mean squared displacement for simulation data (i) All data sets from low mesh size (cyan circles) and high mesh size (green circles) plotted together. These data were fit to the worm-like chain model (Eq. \ref{eq:WLC}) to find the persistence length. (ii) Persistence length ($\mu$m) plotted against mesh size ($\mu$m). The best fit slope is 1.6 $\pm$ 0.2 (iii) Comparison of average of persistence length for low mesh size and high mesh size simulations. All fit parameters can be found in Appendix table \ref{table:FitParameters}. }
\end{figure}
The characteristic persistence length of the trajectories of the quantum dot cargoes was characterized using a mean squared displacement (MSD) as a function of trajectory contour length. Fitting the data this way removed the need to consider time dependence, as described above.  Each MSD was fit to a worm-like chain model (Eq. \ref{eq:WLC}) to determine the persistence length, $L_p$ of the trajectory (Fig. \ref{fig:MSD}Ai,Bi). For both experimental data and simulated data, the MSD was cut off at a countour length of 6 $\mu$m for the fitting in order to compare them.

For the experimental trajectories, the persistence length depended on the mesh size linearly (Fig. \ref{fig:MSD}Aii). The median for low mesh size was 0.44 $\pm$ 0.06 and the median for high mesh size was 0.8 $\pm$ 0.2. The standard deviation for the high mesh size data was large (SD = 0.7 $\pm$ 0.2) compared to the low mesh size data (SD = 0.25 $\pm$ 0.06), which resulted in a probability of 4.2$\%$ that these distributions are the same using the KS Test (Fig. \ref{fig:MSD}Aiii). 

The simulation trajectory data showed the same linear trend with mesh size (Fig. \ref{fig:MSD}Bii), but the data was less spread out for both small and large mesh sizes. The median for low mesh size was 2.8 $\pm$ 0.3 and the median for high mesh size was 4.1 $\pm$ 0.3. The KS test revealed that the probability they are the same distribution is only 0.3$\%$ (Fig. \ref{fig:MSD}Biii).

\section{Conclusion}\label{sec13}
The organization of cytoskeletal filaments is known to affect the motion of motors and cargoes. Here, we took the approach to characterize the network using the mesh size and examine the effects of the mesh size on the run length, displacement, and mean square displacement of the motion through the network. We found that these parameters are sensitive to mesh size, even over the small range in mesh size that we are able to realize in experiments, 1 $\mu$m to 5 $\mu$m. Despite the small range of mesh sizes, there were significant changes in all the trajectory parameters.   

Using experimental trajectories, we deduced the parameters for cargo detachment at intersections and between intersections. We used these parameters to create an analytical theory for the average run length as a function of mesh size, which had similar trends as our experimental data and was sensitive to the probability of detaching at intersections. 

Using the exact same networks extracted from the experimental data and the off rates, we were able to simulate cargo trajectories through the networks. The simulation data had the same trends and similar quantitative results as the experiments for run length, displacement, and mean squared displacement for real cargoes assembled with kinesin motors. We anticipate that future work with different mesh sizes, filament organizations, and motor types will be modeled with the same fundamental principles we uncover here. Specifically, we anticipate that other motors would have different reactions to intersections and composite motor systems would further increase complexity.

\backmatter


\bmhead{Acknowledgments}
We would like to acknowledge the support of the members of the Ross Lab in the Physics Department at Syracuse University. NK was partially supported on funds from Syracuse University, the National Science Foundation grant NSF BIO-2134215 to JLR and National Science Foundation grant DMREF-2118403 to JLR. AG, MG and NS acknowledge support from the National Science Foundation (NSF-DMS-1616926 to AG) and NSF-CREST: Center for Cellular and Biomolecular Machines at UC Merced (NSF-HRD-1547848 and NSF-HRD-2112675 to AG). AG and NS also acknowledge support from the NSF Center for Engineering Mechanobiology grant CMMI-1548571. NS acknowledges Graduate Dean's dissertation fellowship from the University of California, Merced. 

\section*{Declarations}

\begin{itemize}
\item Funding: 
The research leading to these results received funding from Syracuse University and the National Science Foundation grants NSF BIO-2134215 and DMREF-2118403 as well as NSF-DMS-1616926, NSF-HRD-1547848, NSF-HRD-2112675, NSF-CMMI-1548571 and UC Merced Graduate Division. 
Employment: JLR and NK are employed by Syracuse University. NS and AG are employed by University of California, Merced.
\item Data Availability:
Microscopy and simulation data generated for this manuscript are available upon request from the corresponding authors JLR for experimental data and AG for simulation data.
\item Authors' contributions:
NK conceived and design the work, performed data or the acquisition, analysis, and interpretation of data, drafted and edited the manuscript, and is accountable for the work. NS performed simulations and theoretical calculations, analyzed and interpreted data, drafted and edited the manuscript and is accountable for the work. MG performed simulations and  analyzed and interpreted data. AG performed theoretical calculations, interpreted data, drafted and edited the manuscript and is accountable for the work. JLR designed the experimental work, analyzed and interpreted data, drafted and edited the manuscript and is accountable for the work.

\end{itemize}

\begin{appendices}

\section{Experimental Data Appendix}\label{secA1}

In this appendix, we list the parameters of the data for the trajectories for each experimental movie.

\begin{table}[!ht]
\caption{\label{table:NvaluesMeshes} Summary of mesh size data from experimental networks}
\centering
\begin{tabular}{|p{2cm}|p{2cm}|p{2cm}|p{2cm}|p{2cm}|}
\hline
Network name & Number of branches &  Average mesh size & Mesh size standard Error & Number of tracked trajectories \\
\hline
  
HDN001.1 & 566 & 1.675 & 0.025 & 21 \\
HDN001.2 & 620 & 1.73 & 0.029 & 41 \\
HDN001.3 & 423 & 1.839 & 0.039 & 61 \\
HDN001.4 & 601 & 1.676 & 0.024 & 50 \\
HDN002.1 & 581 & 1.678 & 0.026 & 20 \\
HDN002.2 & 536 & 1.784 & 0.035 & 19 \\
HDN002.3 & 418 & 1.882 & 0.039 & 40 \\
HDN002.4 & 469 & 1.765 & 0.033 & 36 \\
HDN005.1 & 1001 & 1.96 & 0.032 & 120 \\
HDN005.2 & 1213 & 1.928 & 0.028 & 115 \\
HDN005.3 & 994 & 1.687 & 0.02 & 185 \\
HDN008.1 & 148 & 2.537 & 0.142 & 24 \\
HDN008.2 & 89 & 3.816 & 0.35 & 40 \\
HDN008.3 & 131 & 3.021 & 0.178 & 56 \\
LDN001.2 & 181 & 2.029 & 0.084 & 34 \\
LDN001.3 & 138 & 2.554 & 0.19 & 32 \\
LDN001.4 & 209 & 1.884 & 0.065 & 41 \\
LDN002.1 & 106 & 2.67 & 0.202 & 41 \\
LDN002.2 & 103 & 2.783 & 0.21 & 53 \\
LDN002.3 & 182 & 2.675 & 0.182 & 44 \\
LDN002.4 & 136 & 2.372 & 0.19 & 49 \\
LDN004.1 & 107 & 3.049 & 0.255 & 69 \\
LDN004.2 & 203 & 2.188 & 0.098 & 71 \\
LDN004.3 & 114 & 3.764 & 0.293 & 117 \\
LDN004.4 & 183 & 2.218 & 0.153 & 108 \\
LDN005.1 & 76 & 3.643 & 0.296 & 36 \\
LDN005.2 & 122 & 3.309 & 0.235 & 41 \\
LDN005.3 & 136 & 3.694 & 0.277 & 56 \\
LDN005.4 & 99 & 3.076 & 0.269 & 42 \\
LDN007.1 & 90 & 3.496 & 0.288 & 48 \\
LDN007.2 & 48 & 4.002 & 0.602 & 28 \\
LDN007.3 & 91 & 3.146 & 0.284 & 52 \\
LDN007.4 & 71 & 4.065 & 0.559 & 63 \\
LDN008.1 & 75 & 2.202 & 0.177 & 12 \\
LDN008.3 & 44 & 4.517 & 0.71 & 37 \\
VHD001.1 & 924 & 1.663 & 0.019 & 82 \\
VHD001.2 & 1003 & 1.687 & 0.019 & 73 \\
VHD001.3 & 858 & 1.696 & 0.022 & 177 \\
VHD002.4 & 1067 & 1.673 & 0.019 & 45 \\
VHD003.3 & 718 & 1.663 & 0.022 & 64 \\
VHD006.3 & 700 & 1.602 & 0.019 & 37 \\
\hline
\end{tabular}

\end{table}
\begin{table}[h]
\centering
\caption{\label{table:FitParameters} The fit parameters for the linear fits to the data as a function of mesh size for all figures, as denoted. }
\begin{tabular}{|p{3cm}|p{1.5cm}|p{1.5cm}|p{1.5cm}|p{1.5cm}|p{1.5cm}|}
\hline
\textbf{Y-axis parameter} & \textbf{Intercept} & \textbf{Slope} & \textbf{R-squared} & \textbf{Chi-squared} & \textbf{Reference to Figure} \\
\hline
Run length - Experiments        & 2.19$\pm$0.27  & 0.52$\pm$0.1  & 0.4  & 11.81 & Fig. 3A(i) \\
Run length - Simulations        & 3.42$\pm$0.26  & 0.42$\pm$0.1  & 0.38 & 6.31  & Fig. 3B(i) \\
Instantaneous speed             & 0.05$\pm$0.01  & 0$\pm$0       & 0.08 & 0.01  & Fig. 5A \\
Average speed                   & 0.88$\pm$0.09  & -0.03$\pm$0.03     & 0.02 & 1.33  & Fig. 5C \\
Average displacement - Experiments                   & 1.21$\pm$0.17   & 0.35$\pm$0.06     & 0.44 & 4.64  & Fig. 6A(i) \\
Average displacement - Simulations                   & 2.17$\pm$0.21  & 0.54$\pm$0.08     & 0.60 & 4.1  & Fig. 6B(i) \\

Persistence length - Experiments & 0.23$\pm$0.3   & 0.22$\pm$0.12 & 0.09 & 13.12 & Fig. 6A(i) \\
Persistence length - Simulations & -2.69$\pm$2.51  & 1.56 $\pm$ 0.19 & 0.22 & 58.9  & Fig. 6B(i) \\
\hline
\end{tabular}
\end{table}

\section{Simulation Appendix}\label{secA1}

\begin{table}[ht]
\centering
\caption{\label{table:NvaluesMeshessimulation} Summary of each mesh size data from simulation results}
\begin{tabular}{|p{2cm}|p{2cm}|p{2cm}|p{2cm}|p{2cm}|}
\hline
Network name & Number of branches &  Average mesh size & Mesh size standard Error & Number of tracked trajectories \\
\hline
H1.1 & 152 & 2.33 & 0.15 & 1000 \\
H1.2 & 154 & 2.40 & 0.14 & 1000 \\
H1.3 & 199 & 2.09 & 0.11 & 1000 \\
H1.4 & 149 & 2.36 & 0.15 & 1000 \\
H2.1 & 235 & 2.20 & 0.09 & 1000 \\
H2.2 & 178 & 2.39 & 0.14 & 1000 \\
H2.3 & 228 & 2.36 & 0.10 & 1000 \\
H5.1 & 230 & 2.34 & 0.13 & 1000 \\
H5.3 & 292 & 2.23 & 0.09 & 1000 \\
H8.1 & 65  & 3.32 & 0.36 & 1000 \\
H8.2 & 60  & 3.33 & 0.38 & 1000 \\
H8.3 & 74  & 3.31 & 0.33 & 1000 \\
H8.4 & 45  & 4.38 & 0.71 & 1000 \\
L1.2 & 27  & 5.58 & 1.10 & 1000 \\
L1.3 & 18  & 4.00 & 1.26 & 1000 \\
L1.4 & 16  & 5.14 & 1.24 & 1000 \\
L2.1 & 43  & 3.21 & 0.28 & 1000 \\
L2.2 & 38  & 3.16 & 0.52 & 1000 \\
L2.3 & 43  & 3.89 & 0.42 & 1000 \\
L2.4 & 13  & 5.20 & 1.32 & 1000 \\
L4.1 & 34  & 4.43 & 0.75 & 1000 \\
L4.2 & 41  & 2.99 & 0.40 & 1000 \\
L4.3 & 54  & 3.84 & 0.51 & 1000 \\
L4.4 & 42  & 4.68 & 1.04 & 1000 \\
L5.1 & 44  & 3.91 & 0.54 & 1000 \\
L5.2 & 25  & 4.57 & 0.76 & 1000 \\
L5.3 & 46  & 4.31 & 0.49 & 1000 \\
L5.4 & 24  & 5.26 & 1.15 & 1000 \\
L7.1 & 34  & 3.61 & 0.60 & 1000 \\
L7.2 & 2   & 7.63 & 6.09 & 1000 \\
L7.3 & 41  & 3.86 & 0.57 & 1000 \\
L7.4 & 27  & 5.15 & 1.27 & 1000 \\
V1.1 & 322 & 2.10 & 0.07 & 1000 \\
V1.2 & 252 & 2.18 & 0.10 & 1000 \\
V1.3 & 266 & 2.09 & 0.08 & 1000 \\
V1.4 & 252 & 2.25 & 0.11 & 1000 \\
\hline
\hline
\end{tabular}
\end{table}

\begin{table}[ht!]
\centering
\caption{\label{table:simulationparameters} Parameters used for running the simulation}
\begin{tabular}{|p{7cm}|p{1cm}|p{2cm}|p{1cm}|}
\hline
\textbf{Description} & \textbf{Units} & \textbf{Variable Name} & \textbf{Value} \\
\hline
Number of cargoes & - & numCargos & 1000 \\
Velocity  & ($\mu m$/s) & v  & 0.1  \\
Cargo radius  & $\mu m$ & cRad & 0.1 \\
Cargo step size & ($\mu m$) & dstep & 0.2 \\
Maximum measuring time & (s) & tMax & 1000  \\
Diffusion constant & ($\mu m$\textsuperscript{2}/s) & D & 1 \\
Rate of detachment from filament & (s\textsuperscript{-1}) & $k_{off}$       & 0.01 \\
Rebinding allowed (1 if allowed, 0 if not allowed) & - & reb\_all & 0 \\
Probability of detachment at intersections   & - & $P_d$ & 0.754 \\
Inverse of the run length without intersections   & $(\mu m^{-1})$ & $\tilde{k_d}$ & 0.1 \\
\hline
\end{tabular}
\end{table}

\begin{figure}
    \centering
    \includegraphics[width=0.5\linewidth]{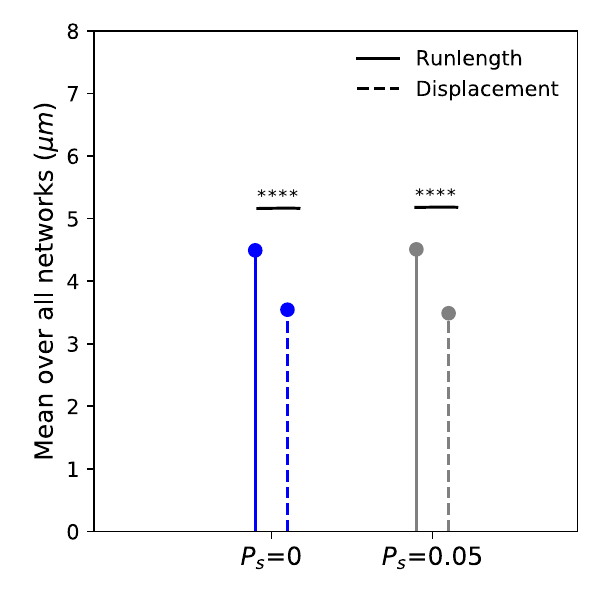}
    \label{fig:simulationfig}
    \caption{Comparison between simulations with different probabilities of switching at intersections ($P_s$). We computed the mean of run lengths and displacements of simulated cargo runs over all extracted networks. No statistically significant difference was seen in mean run length or displacement values between $P_s=0$ and $P_s=0.05$. Statistical significance was determined using students t-test, conventions used: n.s. for $p>0.05$, **** for p$<10^{-4}$.}
    \label{fig:enter-label}
\end{figure}




\end{appendices}


\bibliography{sn-bibliography}

\end{document}